\documentclass{ecai} 
\usepackage{latexsym}
\usepackage{amssymb}
\usepackage{amsmath}
\usepackage{amsthm}
\usepackage{booktabs}
\usepackage{enumitem}
\usepackage{graphicx}
\usepackage{color}
\usepackage{algorithm}
\usepackage{algorithmic}

\newtheorem{theorem}{Theorem}

\newtheorem{definition}{Definition}

\newcommand{\BibTeX}{B\kern-.05em{\sc i\kern-.025em b}\kern-.08em\TeX}

\begin{document}

%%%%%%%%%%%%%%%%%%%%%%%%%%%%%%%%%%%%%%%%%%%%%%%%%%%%%%%%%%%%%%%%%%%%%%%%

\begin{frontmatter}

%%% Use this command to specify your submission number.
%%% In doubleblind mode, it will be printed on the first page.

\paperid{813} 

%%% Use this command to specify the title of your paper.

\title{A Generative Adversarial Graph Neural Network for Synthetic Time Series Data}

%%% Use this combinations of commands to specify all authors of your 
%%% paper. Use \fnms{} and \snm{} to indicate everyone's first names 
%%% and surname. This will help the publisher with indexing the 
%%% proceedings. Please use a reasonable approximation in case your 
%%% name does not neatly split into "first names" and "surname".
%%% Specifying your ORCID digital identifier is optional. 
%%% Use the \thanks{} command to indicate one or more corresponding 
%%% authors and their email address(es). If so desired, you can specify
%%% author contributions using the \footnote{} command.

\author[A, B]{\fnms{Marco}~\snm{Gregnanin}\thanks{Corresponding Author. Email: marco.gregnanin@imtlucca.it}\footnote{Equal contribution.\\ Working Paper}}
\author[B]{\fnms{Johannes}~\snm{De Smedt}\footnotemark}
\author[A]{\fnms{Giorgio}~\snm{Gnecco}\footnotemark} 
\author[C]{\fnms{Maurizio}~\snm{Parton}\footnotemark} 

\address[A]{IMT School for Advanced Studies Lucca}
\address[B]{KU Leuven}
\address[C]{University of Chieti-Pescara}

%%% Use this environment to include an abstract of your paper.

\begin{abstract}
Generating synthetic data for financial time series poses challenges, especially considering their non-stationary nature. Traditional statistical time series models normally assume weak stationarity. However, this assumption can constrain their effectiveness. Deep learning models, particularly Generative Adversarial Networks (GANs), have exhibited considerable potential in emulating complex probability distributions. GANs employ a generator-discriminator framework, where the generator creates data samples, while the discriminator distinguishes real from generated data.\\
In this research, we introduce the Sig-Graph GAN model, which integrates the time-series signature, offering a structured summary of its temporal evolution; the Long Short-Term Memory network, capturing its inherent autoregressive structure; and Graph Neural Networks (GNNs), leveraging  geometric patterns within the time-series data. To employ GNNs optimally, we use the visibility graph algorithm to derive a graph-based representation of the underlying time series.\\
Numerical evaluations demonstrate that the Sig-Graph GAN model outperforms baseline methods in replicating the distribution of logarithmic returns across different stock exchanges. The integration of the graph structure with the autoregressive component effectively captures both geometric and temporal patterns embedded in time-series data.\\
This research advances the field of GAN models for time series by introducing a model capable of leveraging both autoregressive properties and geometric structures for synthetic data generation.
\end{abstract}

\end{frontmatter}

%%%%%%%%%%%%%%%%%%%%%%%%%%%%%%%%%%%%%%%%%%%%%%%%%%%%%%%%%%%%%%%%%%%%%%%%

\section{Introduction}

Across various domains including, among others, healthcare, image and signal processing, economics and finance, environmental science, and energy consumption, the necessity arises for the generation of synthetic data. This need is driven by several factors, such as the scarcity of original data due to privacy concerns, and the requirement for data diversity to enhance model generalization.
However, the generation of synthetic time series data poses a considerable challenge due to their stochastic nature. This holds especially true for the financial case. Two distinct approaches for generating synthetic data are commonly considered: model-based and data-driven. The former involves developing statistical or mathematical models to describe the underlying data-generating process. Classical statistical and financial mathematics models have achieved favorable outcomes in generating synthetic data. %Nonetheless,
Both approaches rely on certain assumptions. Classical statistical models, such as Autoregressive Integrated Moving Average (ARIMA) ~\cite{Box2015}, assume that stock prices exhibit stationarity or at the very least, weakly stationary, implying that both the mean and the auto-covariance of a time series are time-invariant ~\cite{Tsay2005}. Similarly, financial mathematics models, like the Black-Scholes ~\cite{Black1973pricing}, and Merton ~\cite{merton1976option} models, rely on stationarity and homoskedasticity assumptions while also assuming that stock prices reflect all available information ~\cite{fama1970efficient} and follow a random walk ~\cite{lamberton2011introduction}.
In contrast, the data-driven approach seeks to uncover and leverage latent patterns within the data. Machine learning and deep learning models, particularly Generative Adversarial Networks (GANs) ~\cite{goodfellow2020generative}, have exhibited promising results in generating synthetic data. However, existing GAN models in the literature primarily focus on capturing and exploiting the autoregressive component inherent in time series data. For financial time series, this may prove inadequate for capturing non-stationarity and particularly evolving volatility, which exerts a substantial influence on trading strategies owing to its capacity to alter risk assessments associated with stocks ~\cite{buff2002uncertain}. Notably, volatility stands as a pivotal indicator widely employed in the domain of risk management ~\cite{meucci2005risk}. \\
In this paper, we show that geometric patterns play an important role in addressing the generation of synthetic data for a time series. Specifically, transforming time series from a Euclidean to a non-Euclidean space using a graph-based approach can significantly enhance our understanding and analysis of complex financial time series behavior as the examination of the graph's node degree distribution can provide insights into whether the series showcases periodic, random, or fractal characteristics ~\cite{lacasa2008time}, which is particularly relevant in financial time series given the presence of fractal behavior in stock prices ~\cite{mandelbrot2013fractals}. Introducing fractal concepts into financial modeling can lead to a more accurate depiction of market dynamics. Fractals possess the capacity to encapsulate the inherent self-similarity and scaling properties that are intrinsic to financial markets ~\cite{peters1994fractal,evertsz1995fractal}. Furthermore, the adoption of a graph-based representation for the data liberates us from the constraint of assuming stationarity within the time series. The graph representation facilitates the exploration of structural dynamics intrinsic to the time series, providing a means to apprehend relationships and dependencies among the data points.
Graph Neural Networks (GNNs) ~\cite{scarselli2008graph} possess the capability to capture and model the geometric patterns intrinsic to graph data. Analogous to fractals, GNNs can capture local structural information within a graph by iteratively aggregating data from neighboring nodes. This iterative process empowers GNNs to acquire and generalize patterns at varying scales, thereby effectively capturing the intrinsic geometric attributes of the graph akin to fractals embodying self-similarity. Furthermore, we explore the application of the time series signature ~\cite{lyons1998differential,lyons2014feature}, a concept derived from path theory, which provides a universal description of time series. In essence, the signature can be viewed as analogous to the Moment Generating Function (MGF), which is significant for comparing random variable distributions as it encodes all distribution moments into a single function, uniquely characterizing the distribution itself ~\cite{resnick2019probability}. \\
To address the challenge of generating synthetic financial time series data, we propose an innovative GAN model that adeptly harnesses both the autoregressive nature of time series and the underlying geometric properties. This is achieved by integrating two techniques, the visibility graph and the time series signature
%by combining the visibility graph and the time series Signature, two techniques   
that are traditionally applied in the domains of Complex Network Analysis, and Stochastic Analysis, %(namely, the visibility graph and the time series Signature), 
respectively, into %within the context of a generative adversarial model}
the framework of a generative adversarial model. Our contributions encompass:
\begin{itemize}
    \item[1)] Introducing a novel GAN model that integrates dynamic GNNs with the Long Short-Term Memory (LSTM) network ~\cite{hochreiter1997long}; 
    \item[2)] Proposing two custom loss functions based on the signature, derived specifically for the Sig-Graph GAN model, and based on the Mean Square Error (MSE) and Kullback-Leibler Divergence (KLD) metrics. These functions are proposed in order to capture both pointwise discrepancies and probability distribution function (PDF) disparities between real and generated data;
    \item[3)] Conducting a comprehensive evaluation of our approach across diverse stock exchanges, showcasing performance enhancements compared to benchmark models. 
\end{itemize}
The subsequent sections of this paper are organized as follows: Section \ref{Section_2} reviews relevant statistical and GAN models used for generating time series data; Section \ref{Section_3} defines the time series signature, the visibility graph, and the GNN model; Section \ref{Section_4} precisely defines the problem under consideration; Section \ref{Section_5} elucidates the architecture of our proposed model; Section \ref{Section_6} provides a comparative analysis of our model against benchmark models; and finally, Section \ref{Section_7} concludes the paper.

%%%%%%%%%%%%%%%%%%%%%%%%%%%%%%%%%%%%%%%%%%%%%%%%%%%%%%%%%%%%%%%%%%%%%%%%

\section{Related Works}
\label{Section_2}
 
The generation of financial time series data is a crucial task due to their limited availability. %Specifically, 
While daily data is openly accessible, high-frequency data are typically proprietary. However, relying solely on accessible daily data might be insufficient, leading to potential biases and overfitting in models ~\cite{assefa2020generating}.\\
Generating financial time series data can be accomplished through either model-based or data-driven methodologies. The former involves the utilization of statistical and mathematical models. Prominent statistical techniques encompass the Autoregressive Integrated Moving Average (ARIMA), Autoregressive Conditional Heteroskedasticity (ARCH), and Generalized AutoRegressive Conditional Heteroskedasticity (GARCH) models ~\cite{brockwell2002introduction}. These models vary in their treatment of volatility; ARIMA assumes constant volatility, while others permit varying variances. Notably, ARCH and GARCH have been employed to model and forecast the volatility of stock exchanges such as Nasdaq and Dow Jones ~\cite{engle2012arch}. Mathematical models predominantly rely on geometric Brownian motion. %~\cite{bachelier1900theorie}. 
The Black-Scholes-Merton model ~\cite{Black1973pricing} %,merton1974pricing}
utilizes geometric Brownian motion for studying asset price paths and pricing European options. ~\cite{merton1976option} and ~\cite{kou2002jump} introduced discontinuity to capture fat-tailed distributions, and ~\cite{heston1993closed} modeled stochastic volatility. \\%More generalized approaches adopt Levy processes for stock price paths ~\cite{tankov2003financial}. \\
Conversely, data-driven approaches leverage observed data for pattern extraction. Monte Carlo simulations ~\cite{wang2012monte} and Bootstrapping ~\cite{ruiz2002bootstrapping} are common for generating stock prices and return distributions. Nonetheless, these methods inherit assumptions from the model-driven approaches, potentially affecting their results. Machine and deep learning models are also employed, with GANs ~\cite{goodfellow2020generative} achieving success in various domains. SeqGAN ~\cite{yu2017seqgan} was introduced for the generation of sequential discrete time series, exemplified in domains like music and Speech-language. C-RNN-GAN ~\cite{mogren2016crnngan} employed Recurrent Neural Networks (RNNs) and bidirectional RNNs to generate music. TimeGAN ~\cite{Yoon2019TimeGAN} was developed to maintain temporal dynamics within time series data, as observed in domains like energy and stock prices. Concentrating exclusively on the generation of financial time series, we have Quant GANs ~\cite{wiese2020quant} that utilize Temporal Convolutional Networks (TCNs) for capturing long-term dependencies in the financial time series. %generation. 
The Restricted Boltzmann Machine (RBM) generates synthetic market data by utilizing stochastic binary activation units for the visible and hidden layers ~\cite{kondratyev2019market}. To enhance GAN performance, Wasserstein loss was introduced ~\cite{de2019enriching} in order to measure the distance between two probability distributions. The integration of GANs with the time series signature was proposed by ~\cite{ni2020conditional},  who introduced the Signature Wasserstein-1 metric (Sig-$W_1$) as loss function. ~\cite{de2022tackling} extended Sig-$W_1$ for multi-stock price generation. DAT-CGAN ~\cite{sun2023decisionaware} was proposed into the multi-period asset allocation problem. 
However, all these methods focus solely on the autoregressive nature of time series data when generating synthetic sequences, neglecting the geometric patterns.%without considering the geometric pattern in the data}.

%%%%%%%%%%%%%%%%%%%%%%%%%%%%%%%%%%%%%%%%%%%%%%%%%%%%%%%%%%%%%%%%%%%%%%%%

\section{Preliminaries}
\label{Section_3}

This section offers the mathematical underpinnings of the methodology. We begin by defining the path signature, followed by the visibility graph algorithm used for the graphical representation of univariate time series. Lastly, we briefly outline the generic mathematical foundations of GNN, and GAN models.
\subsection{Signature}
The signature is a mathematical concept derived from the field of path theory, which provides a comprehensive and structured representation of the temporal evolution of a time series. In accordance with the notation presented in ~\cite{ni2020conditional}, for the sake of convenience, our discourse shall be limited to the space of continuous functions mapping from a compact time interval $J:=[a,b]$ to $\mathbb{R}^d$ with finite $p$-variation and starting from the origin, denoted by $C^{p}_0(J, \mathbb{R}^d)$.\\
Let $\mathrm{T}((\mathbb{R}^{d})):=\oplus_{k=0}^{\infty}(\mathbb{R}^{d})^{\otimes k}$ be a tensor algebra space. This space encapsulates the signatures of $\mathbb{R}^{d}$-valued paths, allowing for their comprehensive representation. 
Moreover, let $S=\{s_1, s_2, \dots, s_T\}$ denote a discrete time series. Subsequently, it becomes necessary to convert the time series into a continuous path using either the lead-lag transformation or the time-join transformation method ~\cite{levin2013learning}. Let $L$ denote the continuous path derived through the lead-lag transformation, which is employed within our model. Within this context, the signature $\mathcal{S}$ and the truncated signature at level $M$, denoted as $\mathcal{S}_M$, can be defined as follows:
\begin{definition}[Signature and Truncated Signature]
Let \(L \in C^{p}_0(J, \mathbb{R}^d)\) be a path. The signature \(\mathcal{S}\) of the path \(L\) is defined as:
\begin{equation}
    \mathcal{S} = (1, L_{J}^{1}, \dots, L_{J}^{k}, \dots ) \in \mathrm{T}(\mathbb{R}^{d})\,, \label{eq:signature}
\end{equation} 
where \(L_{J}^{k}=\int_{t_1\ < t_2 < \dots t_k, t_1, \dots t_k \in J}{dL_{t_{1}}\otimes \dots \otimes dL_{t_{k}}}\) are called iterated integrals. \\
The truncated signature of degree \(M\) is defined as:
\begin{equation}
    \mathcal{S}_M = (1, L_{J}^{1}, \dots, L_{J}^{M} )\,. \label{eq:truncated_signature}
\end{equation} 
\end{definition}
The signature of a path offers a hierarchical interpretation: lower-order components capture the general path attributes, while higher-order terms unravel intricate local characteristics. Notably, the signature is invariant to reparameterization, preserving the integral values despite transformations at different time instances. It is also translation invariant and adheres to concatenation properties ~\cite{chen1958integration}. The truncated signature retains the first $\frac{d^{M+1}-1}{d-1}$ iterated integrals, with $M$ representing the truncation degree, and $d$ the dimension of the path. The factorial decay of the neglected iterated integrals %, $\mathcal{O}(\frac{1}{M!})$,
ensures that minimal information loss occurs in truncating the signature \(\mathcal{S}\) ~\cite{lemercier2021distribution}.\\ 
Given two stochastic processes, denoted as $A$ and $B$, defined on a probability space $(\Omega, \mathbb{P}, \mathcal{F})$, suppose that equation (\ref{eq:signature}) is satisfied almost surely for both $A$ and $B$. Additionally, assume that the expected values of $\mathcal{S}(A)$ and $\mathcal{S}(B)$ are finite. We have the following theorem ~\cite{lyons2015expected}:
\begin{theorem}[Expected Signature] Let \(A\) and \(B\) be two \(C_{0}^{1}(J,\mathbb{R}^{d})\)-valued random variables. If \(\mathbb{E}[\mathcal{S}(A)]=\mathbb{E}[\mathcal{S}(B)]\), and \(\mathbb{E}[\mathcal{S}(A)]\) has infinite radius of convergence, then \(A \overset{d}{=} B\), i.e., $A$ and $B$ are equal in distribution.  
\end{theorem}
While a path's signature uniquely defines its trajectory ~\cite{lyons1998differential}, the expected signatures uniquely ascertain the paths' distributions, akin to the role of the moment generating function ~\cite{chevyrev2016characteristic}. \\
For further comprehensive definitions and rigorous formulations, we refer to ~\cite{lyons2014rough,chevyrev2016primer,levin2013learning}.

\subsection{Network Theory and Visibility Graph}
Let $G=(V,E)$ be a graph, where $V=\{v_1, \dots, v_n\}$ represents the set of nodes, and $E\subseteq V \times V$ constitutes the set of edges. The associated graph $G$'s adjacency matrix $A$ is an $n\times n$ matrix whose entries are 1 or 0, indicating whether nodes $u$ and $v$ are connected by an edge, i.e., $A(u,v)=1$ if and only if $(u,v) \in E$.\\
In this study, we establish a dynamic graph through a snapshot-based representation ~\cite{wang2019time}. A dynamic graph constitutes a sequence of static graph snapshots in discrete time, incorporating a set $\mathcal{T}=\{t_1, t_2, \dots, t_T\}$ encompassing $T=|\mathcal{T}|$ time steps, i.e., $\mathcal{G} = \{G_{t_i}\}_{i=1}^T$, where $G_{t_i} = (V_{t_i}, E_{t_i})$ is the snapshot taken at time $t_i$. \\ %with $V_t \subseteq V$ and $E_t \subseteq E$. \\
We use the visibility graph algorithm ~\cite{lacasa2008time} to generate a graph from a univariate time series, which establishes connections between data points if there exists a clear line of sight between them. This approach involves depicting time series values as vertical bins, with edges formed between each bar and the visible bars from the top, devoid of obstructions. Each data point becomes a node, while edges represent visibility connections. Given a univariate time series $S=\{s_1, s_2, \dots, s_T\}$ and associated discrete time stamps $\mathcal{T}=\{t_1, t_2, \dots, t_T\}$, the visibility criterion is defined as follows: any two data points $(s_i, t_i)$ and $(s_j, t_j)$ link in the graph if every other data point $(s_k, t_k)$ in a suitable set satisfies: $s_k < s_j + (s_j - s_i) \frac{t_j - t_k}{t_j - t_i}$.
%\begin{equation*}
%s_k < s_j + (s_j - s_i) \frac{t_j - t_k}{t_j - t_i}\,.
%\label{eq: visibility_graph}
%\end{equation*}
In time series data, the visibility graph approach captures non-linear dependencies and patterns ~\cite{lacasa2008time}. Furthermore, despite affine transformations, the resulting time series-based graph stays connected and invariant ~\cite{stephen2015visibility}. This approach allows one to create either undirected or directed graphs starting from a time series. For the former, the visibility condition considers all potential $(s_i, t_i), i=1,\dots,T$ value combinations. In contrast, for directed graphs, the visibility condition is applied under the ``from left to right'' principle, where $(s_i, t_i)$ is compared to values with time stamps greater than $t_i$, i.e., $\forall (s_j, t_j), j=i+1, \dots, T$.\\ 
\begin{figure}[t]
\begin{center}
\includegraphics[width=0.53\textwidth, height=0.14\textheight, trim={1.5cm 0 0 0}]{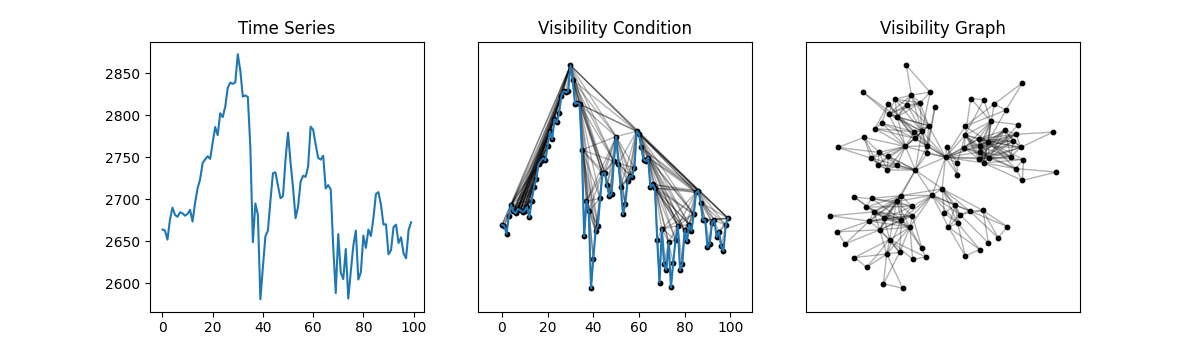}
\end{center}
\caption{Visibility graph algorithm applied to the Standard and Poor's 500 closing price from 2017/12/12 to 2018/05/07.} \label{fig:visibility_graph}
\end{figure}
Figure \ref{fig:visibility_graph} demonstrates the application of the visibility graph algorithm to a univariate time series, computed using the ``Time series to visibility graphs'' (ts2vg) Python package ~\cite{Bergillos2020visibility}. The first plot illustrates closing price trajectories, the middle plot depicts the visibility condition and link formation, and the final plot presents the corresponding undirected graph. %The visibility graph algorithm preserves multiple time series properties in the graph representation. For instance, periodic time series translate to regular graphs, where each vertex possesses an identical number of neighbors. Conversely, random time series manifest as random graphs 
The visibility graph transformation preserves structural properties of the time series: periodic time series are transformed into regular graphs, random time series correspond to random graphs, and fractal time series are represented as small-world graphs ~\cite{lacasa2008time}. Hence, the visibility graph serves as an effective means for representing non-stationary time series, ensuring that no information pertaining to temporal aspects is lost in the process, and capturing the inherent patterns and trends present in a time series, such as peaks, valleys, and overall trends.

\subsection{Graph Neural Networks}
Graph Neural Networks (GNNs) are sophisticated deep learning models engineered to derive node embeddings for graphs. Built upon a message-passing paradigm, GNNs facilitate the dissemination of information among graph nodes ~\cite{gilmer2017neural,xu2018representation}. Following the formalism presented in ~\cite{you2022roland}, we establish the $l$-th layer within a GNN as follows:
\begin{align}
m^{(l)}_{u\rightarrow v} &= \text{MSG}^{(l)}(h_{u}^{(l-1)}, h_{v}^{(l-1)})\,, \nonumber \\
h_{v}^{(l)} &= \text{AGG}^{(l)}({m^{(l)}_{u\rightarrow v} | u \in N(v) }, h{_v}^{(l-1)})\,. \label{eq:message_passing}
\end{align}
Here, $h_{v}^{(l)}$ denotes the embedding of node $v$, initialized with the feature value of node $v$ as $h_{v}^{(0)}=x_v$. The function $\text{MSG}(\cdot)$ represents the message function, yielding the message embedding $m^{(l)}$ at the $l$-th layer. $N(v)$ signifies the set of neighboring nodes for node $v$, and $\text{AGG}(\cdot)$ symbolizes the aggregation function, elaborated later. The role of the message function is to facilitate a seamless exchange of information among nodes and their neighbors. A prevalent practice involves employing a linear function to compute the message content. Conversely, the aggregation function amalgamates received messages from neighboring nodes, fashioning a coherent representation for the recipient node. The aggregation function needs to remain permutation invariant and generate consistent node representations. Common aggregation functions encompass mean, sum, and maximum functions. The label ``$l$-th layer'' encapsulates information originating from nodes situated at a distance of $l$ hops away, providing information diffusion across successive GNN layers.

\subsection{Generative Adversarial Networks}
Generative Adversarial Networks (GANs) ~\cite{goodfellow2020generative} represent a class of machine learning models structured around a strategic game between two agents, the generator and the discriminator. The fundamental objective of a GAN is to achieve a probability distribution for the generative model that closely approximates the distribution of the actual data $\mathbf{x}$, i.e., $\mathbb{P}_{\text{model}}(\mathbf{\tilde{x}}) \sim \mathbb{P}_{\text{data}}(\mathbf{x})$ ~\cite{goodfellow2020generative}.\\
In this study, the generator, denoted as \textit{Gen}, takes random noise $\mathbf{z}$, and an adjacency matrix $A$ as input and leverages a vector of learnable parameters $\mathbf{\theta}$ to generate $\mathbf{\tilde{x}}$, yielding $\mathbf{\tilde{x}}=\textit{Gen}(\mathbf{z}, \mathbf{\theta}, A)$. %model $\mathbb{P}_{\text{model}}$, yielding $\mathbb{P}_{\text{model}}(X)=\textit{Gen}(\mathbf{z}, \mathbf{\theta})$. 
Here, $\textit{Gen}(\cdot)$ is the generator function, while $A$ is the adjacency matrix associated to the graph representation of the true data, and $\mathbf{z}$ originates from a probability distribution denoted as $\mathbb{P}_{\mathbf{z}}(\mathbf{z})$ — typically assumed to be a normal or uniform distribution. Conversely, the discriminator, labeled as \textit{Dis}, seeks to distinguish between genuine data and synthetic data. Specifically, the discriminator takes real data and the adjacency matrix $A$ as input, and produces the probability of it being authentic with a vector of learnable parameters $\mathbf{\eta}$, i.e., $\mathbb{P}_{\text{real}}(\mathbf{x})=\textit{Dis}(\mathbf{x}, \mathbf{\eta},A)$, where $\textit{Dis}(\cdot)$ represents the discriminator function. Both networks endeavor to minimize their respective costs, participating in a zero-sum game. Consequently, the training of the GAN model follows a minimizing-maximizing approach for the value function $V(\textit{Gen}, \textit{Dis})$:
\begin{align*}
    \min_{\textit{Gen}} \max_{\textit{Dis}} V(\textit{Gen}, \textit{Dis}) &= \mathbb{E}_{\mathbf{x}\sim \mathbb{P}_{\text{data}}(\mathbf{x})}[\log \textit{Dis}(\mathbf{x},A)] \nonumber \\ 
    &+ \mathbb{E}_{\mathbf{z}\sim \mathbb{P}_{\mathbf{z}}(\mathbf{z})}[\log (1 -  \textit{Dis}(\textit{Gen}(\mathbf{z},A),A))]\,. \label{eq:GAN_minmax_function}
\end{align*}
This game between the generator and the discriminator leads to the iterative improvement of the generator's capability to produce realistic data and the discriminator's ability to distinguish between real and synthetic data. It is noteworthy that the training of the discriminator's and generator's parameters, denoted as $\eta$ and $\theta$ respectively, is accomplished by alternating the computation of their gradients and updating their corresponding parameters. 

%%%%%%%%%%%%%%%%%%%%%%%%%%%%%%%%%%%%%%%%%%%%%%%%%%%%%%%%%%%%%%%%%%%%%%%%
\section{Problem Formulation}
\label{Section_4}
Consider a univariate time series $S=\{s_1, s_2, \dots, s_T\}$ extending up to time $T$. The objective of this research is to determine a function $f(\cdot)$, given a random variable $Z$ and an adjacency matrix A associated to the graph-based representation $G$ of the original time series $S$, that is able to generate synthetic data $S'$. The synthetic data should closely resemble the statistical characteristics, temporal dependencies, and geometric patterns observed in the original time series $S$. Therefore, we want to find that $\{s_1, \dots, s_T\}\overset{d}{\simeq} \{s'_1, \dots s'_T\}$, where $S' = f(Z, A)$.\\
Hence, in this study, our objective is to train a GAN model to enable the generator component, represented by $\textit{Gen}(\cdot)$, to %produce a random variable possessing the same distribution as the target time series 
generate synthetic time series data that exhibits as much as possible the statistical properties of the target time series $S$ at time $t$. This generation is accomplished using: a noise random variable $\mathbf{z}$ sampled from the probability distribution $\mathbb{P}_{z}$, the vector of parameters $\mathbf{\theta}$ subject to training, and the adjacency matrix $A_t$ associated to the graph $G_t$ corresponding to the temporal series $S$ evaluated at time $t$, with $t\leq T$. The model is formulated as follows:
\begin{equation*}
S_t'=\textit{Gen}(Z_t, \mathbf{\theta}, A_t) \overset{d}{\simeq} S_t\,. \label{eq:generator_definition}
\end{equation*}
Here, $A_t$ signifies the graphical representation of the time series $S$, incorporating the most recent observation and past $m$ observations until time $t$. Consequently, the set of observations can be expressed as $S_t = \{s_{t-m}, \dots, s_{t}\}$, comprising $m+1$ elements. The random noise $Z_t$ is constructed by sampling $m+1$ observations from a Gaussian distribution with mean and variance of $0$ and $1$, respectively. This procedure is repeated $F$ times, and yields a matrix of noise random vectors, denoted as $Z_t \in \mathbb{R}^{\Tilde{T}\times F}\sim N(0,1)$, where $\Tilde{T}$ conveniently represents $m+1$.

\section{Proposed Sig-Graph GAN Framework}
\label{Section_5}
\begin{figure*}[t]
\centering
\includegraphics[width=0.67\textwidth]{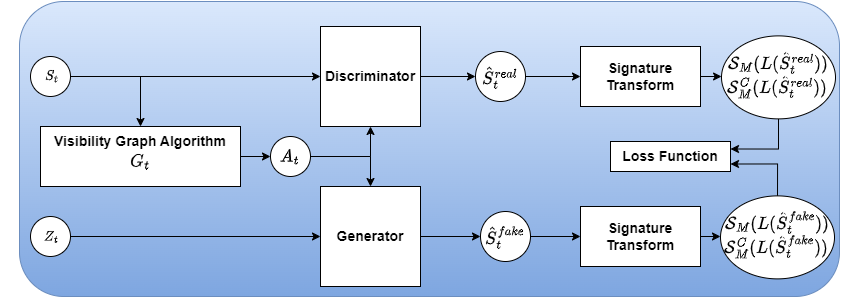} % Reduce the figure size so that it is slightly narrower than the column.
\caption{Proposed Generative Adversarial Networks Architecture.}
\label{fig:prposed_model}
\end{figure*}

\begin{figure}[t]
\centering
\includegraphics[width=\columnwidth]{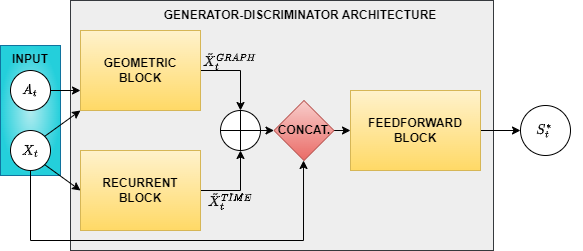}
\caption{Network structure for the Discriminator and Generator Agents.}
\label{fig: network_structure} 
\end{figure}

Figure \ref{fig:prposed_model} illustrates the architecture of the proposed model. In alignment with the principles of the Quant GAN model ~\cite{wiese2020quant}, we adopt an identical network configuration for both the generator and the discriminator, denoted as $\textit{Dis}(\cdot)$. The initialization involves the construction of the time series $S_t = \{s_{t-m}, \dots, s_{t}\}$ and the random noise matrix $Z_{t} \in \mathbb{R}^{\Tilde{T}\times F}$. Subsequently, the corresponding graph $G_t$ associated with the time series is formed utilizing the visibility graph algorithm. The choice between an undirected or directed graph is considered a hyperparameter to optimize. 
Regardless, the adjacency matrix $A_t$ associated to the graph $G_t$ maintains dimensions of $\Tilde{T} \times \Tilde{T}$, with each node corresponding to a specific time observation. The generator/discriminator functions are structured as follows:
\begin{align*}
    & \textit{Dis}(S_t, \mathbf{\eta}, A_t)\,, %\nonumber \\
    & \textit{Gen}(Z_t, \mathbf{\theta}, A_t)\,.
\end{align*}
Here, $\mathbf{\eta}$ represents the vector of learnable parameters for the discriminator, and
$S_{t}\in \mathbb{R}^{\Tilde{T}\times 1}$. The network structure, visualized in Figure \ref{fig: network_structure}, combines a Geometric Block, a Recurrent Block, and a Feedforward Block, for fortifying the discriminator and generator components. For simplicity, we denote the input for the generator and discriminator agents as $Z_t$ and $S_t$, respectively, which can be represented as $X_{t} \in \mathbb{R}^{\Tilde{T}\times F}$, where $F$ is set to $1$ for the discriminator.

\subsection{Recurrent Block}
The Recurrent Block takes the feature matrix $X_t \in \mathbb{R}^{\Tilde{T} \times F}$ as input and produces the output $\Tilde{X}^{TIME}_t \in \mathbb{R}^{\Tilde{T} \times F}$. Its objective is to analyze and leverage temporal patterns within the data. Algorithm \ref{alg:recurrent_block} outlines the operations of this component.

\begin{algorithm}[ht]
   \caption{Recurrent Block}
   \label{alg:recurrent_block}
\begin{algorithmic}
   \STATE {\bfseries Input:} feature matrix $X_t$
   \STATE {\bfseries Output:} temporal pattern $\Tilde{X}^{TIME}_t$
   \STATE $R^{(0)}_t = X_t$ \qquad %\COMMENT{Initialization of the Hidden Layer}
   \FOR{$l=1$ {\bfseries to} $L$}
   \STATE $R^{(l)}_t = \text{LSTM}^{(l)}(R^{(l-1)}_t)$ %\qquad %\COMMENT{See Equation (\ref{eq:g_function_definition})}
   \ENDFOR
   \STATE $\Tilde{X}^{TIME}_t = \text{FC}(R^{(L)}_t)$ %\qquad \COMMENT{See Equation (\ref{eq:fully_connected_layer})}
\end{algorithmic}
\end{algorithm}

The process involves initializing a hidden layer, denoted as $R_{t}^{(0)}$, with the feature matrix. Subsequently, based on the number of layers $L$, the $l$-th hidden layer of a LSTM neural networks model - $R^{(l)}_t = \text{LSTM}^{(l)}(R^{(l-1)}_t)$ - is applied. The LSTM model is consider in this research for its ability to capture temporal and long-term patterns. Finally, to obtain the output of the Recurrent Block, a fully connected layer denoted as $\text{FC}(\cdot)$, is applied, defined by:
\begin{equation}
    \Tilde{X}^{TIME}_t  = W_t R^{(L)}_t + b_t  \,,\label{eq:fully_connected_layer}
\end{equation}
where $W_t$ is the weight matrix, $b$ is a bias vector, %$\phi (\cdot)$ denotes the activation function, 
and $\Tilde{X}^{TIME}_t \in \mathbb{R}^{\Tilde{T} \times F}$ as output, where $F$ equates to $1$ for the discriminator.
%The recurrent block processes input denoted as $Z_t$ for the generator and $S_t$ for the discriminator agent. For simplification, we represent this input as $X_{t} \in \mathbb{R}^{\Tilde{T}\times F}$, where $F$ is set to $1$ for the discriminator. Subsequently, the input traverses through LSTM layers, serving to capture temporal and long-term patterns. The number of layers, neurons, and the recurrent dropout rate constitute hyperparameters subject to optimization. As exemplified in Figure \ref{fig: network_structure}, a two-layer LSTM is demonstrated. Specifically, the LSTM receives $X_{t} \in \mathbb{R}^{\Tilde{T}\times F}$ as input and yields $O_{t} \in \mathbb{R}^{\Tilde{T}\times K_1}$ as output. Here, $K_1$ signifies the number of neurons incorporated. Intervening between the two LSTM blocks is a recurrent dropout layer. Subsequent to this, a fully connected layer concludes the temporal block, producing $\hat{X}^{1}_{t} \in \mathbb{R}^{\Tilde{T} \times F}$ as output, where $F$ equates to $1$ for the discriminator.

\subsection{Geometric Block}
The Geometric Block takes as input the adjacency matrix $A_t \in \mathbb{R}^{\Tilde{T} \times \Tilde{T}}$, derived using the visibility graph on the set $S_t$ of the last $m$ observations of the original time series $S$, and the feature matrix $X_t \in \mathbb{R}^{\Tilde{T} \times F}$. The output of this block represents the geometric patterns and is denoted as $\Tilde{X}^{GRAPH}_t \in \mathbb{R}^{\Tilde{T} \times F}$. Algorithm \ref{alg:geometric_block} outlines the operations of this component.

\begin{algorithm}[ht]
   \caption{Geometric Block}
   \label{alg:geometric_block}
\begin{algorithmic}
   \STATE {\bfseries Input:} adjacency matrix $A_t$, feature matrix $X_t$
   \STATE {\bfseries Output:} geometric pattern $\Tilde{X}^{GRAPH}_t$
   \STATE $H^{(0)}_t = X_t$ 
   \FOR{$l=1$ {\bfseries to} $L^{'}$}
   \STATE $H^{(l)}_t = \text{GNN}^{(l)}(H^{(l-1)}_t)$ %\qquad \COMMENT{See Equation (\ref{eq:gnn_equation})}
   \ENDFOR
   \STATE $\Tilde{X} = \text{LSTM}(H^{(L^{'})}_t)$
   \STATE $\Tilde{X}^{GRAPH}_t = \text{FC}(\Tilde{X})$ 
\end{algorithmic}
\end{algorithm}

The process begins by initializing the hidden layer, denoted as $H^{(0)}_t$, with the feature matrix. Depending on the number of layers $L^{'}$, the $l$-th layer of a GNN is applied, denoted as $H^{(l)}_t = \text{GNN}^{(l)}(H^{(l-1)}_t)$. This can be defined using the message-passing paradigm defined in Equation (\ref{eq:message_passing}). Note that changing the definition of the message and aggregation function leads to different types of GNN models. In this research, we decide to consider only the Graph Convolutional Networks (GCNs) \cite{kipf2016semi}, and the $l$-th layer can be defined as:
\begin{equation*}
    H^{(l)}_t = \rho\left(\Tilde{D}_{t}^{-1/2}\Tilde{A}_{t} \Tilde{D}_{t}^{-1/2}H^{(l-1)}_t\Theta\right)\,,
\end{equation*}
where $\rho(\cdot)$ is the hyperbolic tangent activation function, $\Theta$ is the weight matrix, $\Tilde{A}_{t} = A_t + I$, $I$ is the identity matrix, $\Tilde{D}_{t}$ is the degree matrix computed on $\Tilde{A}_{t}$. Before obtaining the output of the Geometric Block, we pass $H^{(L^{'})}_t$ through an LSTM layer to process the spatial patterns uncovered by the GNN layers, and through a FC layer as defined in Equation (\ref{eq:fully_connected_layer}).

\subsection{Feedforward Block} 

The inputs of the Feedforward Block are the temporal patterns $\Tilde{X}^{TIME}_t \in \mathbb{R}^{\Tilde{T} \times F}$, obtained from the Recurrent Block, the geometric patterns $\Tilde{X}^{GRAPH}_t \in \mathbb{R}^{\Tilde{T} \times F}$, computed by the Geometric Block, and the feature matrix $X_t \in \mathbb{R}^{\Tilde{T} \times F}$. Algorithm \ref{alg:ffn_block} outlines the operations of this component. 

\begin{algorithm}[ht]
   \caption{Feedforward Block}
   \label{alg:ffn_block}
\begin{algorithmic}
   \STATE {\bfseries Input:} temporal pattern $\Tilde{X}^{TIME}_t$, geometric pattern $\Tilde{X}^{GRAPH}_t$, feature matrix $X_t$
   \STATE {\bfseries Output:} prediction $\hat{S}_{t}^{*}$
   \STATE $\Tilde{X}^{TOT}_t = \Tilde{X}^{TIME}_t + \Tilde{X}^{GRAPH}_t$ 
   \IF{Skip Layer True}
    \STATE $\Tilde{X}^{TOT}_t = \text{CONCAT}(\Tilde{X}^{TOT}_t,X_t )$
   \ENDIF
   \STATE $Y^{(0)}_t = \Tilde{X}^{TOT}_t$
   \FOR{$l=1$ {\bfseries to} $L^{''}$}
   \STATE $Y^{(l)}_t = \phi(\text{FC}^{(l)}(Y^{(l-1)}_t))$ %\qquad \COMMENT{See Equation (\ref{eq:fully_connected_layer})}
   \ENDFOR
   \STATE $\hat{S}_{t}^{*} = Y^{(L^{''})}_t$ 
\end{algorithmic}
\end{algorithm}
The process begins by summing up the two patterns, and the result is denoted as $\Tilde{X}^{TOT}_t$. Then, if a skip layer is considered, $\Tilde{X}^{TOT}_t$ is concatenated with the feature matrix $X_t$. Finally, the hidden layer, denoted as $Y^{(0)}_t$ is initialized with $\Tilde{X}^{TOT}_t$. Before obtaining the prediction vector, denoted as $\hat{S}^{*}_{t}\in \mathbb{R}^{\Tilde{T} \times 1}$, the $l$-th layer of a fully connected model is applied, with the number of layers denoted as $L^{''}$. In this research, we decided to consider three, i.e. $L^{''}=3$, fully connected layers, as defined in Equation (\ref{eq:fully_connected_layer}), with $128$, $64$, and $1$ units. Intermediate to these layers lies the Parametric Rectified Linear Unit (PReLU) ~\cite{he2015delving} activation function, corresponding to $\phi(\cdot)$ in the algorithm \ref{alg:ffn_block}, defined as $f(x) = \max (0, x) + \alpha \min (0, x)$ with $\alpha = 0.25$, serving to capture final non-linear patterns.
%Within the Feedforward Block, the outputs from both the recurrent and geometric blocks, $\hat{X}^{1}_{t} \in \mathbb{R}^{\Tilde{T} \times F}$ and $\hat{X}^{2}_{t} \in \mathbb{R}^{\Tilde{T} \times F}$ respectively, are processed. The initial step involves summing the two outputs, followed by concatenation with the initial input ${X}_{t} \in \mathbb{R}^{\Tilde{T} \times F}$. Subsequently, three fully connected layers with unit counts of $128$, $64$, and $1$ are applied. Intermediate to these layers lies the Parametric Rectified Linear Unit (PReLU) ~\cite{he2015delving} activation function, defined as $f(x) = \max (0, x) + \alpha \min (0, x)$ with $\alpha = 0.25$, serving to capture final non-linear patterns. The final output of the network is denoted as $\hat{S}^{*}_t$, wherein $*$ is replaced with ``\textit{real}'' for the discriminator and ``\textit{fake}'' for the generator.\\

\subsection{Loss Functions} %or Custom Loss Function? or Proposed Loss Function? 
Prior to computing the loss function, we subject the output of the discriminator and generator agent $\hat{S}^{*}_t$, - wherein $*$ is replaced with ``\textit{real}'' for the discriminator and ``\textit{fake}'' for the generator - , to the lead-lag transformation, denoted with $L(\cdot)$.
To this end, we apply the truncated signature, as defined in Equation (\ref{eq:truncated_signature}), with a truncation level degree set at $5$. In further pursuit of comprehensive insight, we also perform a cumulative summation on $\hat{S}^{*}_t$, subsequently leading to the computation of a cumulative truncated signature ~\cite{chevyrev2016primer}, which retains path dependencies, and mirroring the temporal progression of information accumulation. We compute the signature using the ``iisignature'' Python packages ~\cite{kidger2020signatory}.\\
The two custom loss functions adopted are Mean Square Error (MSE) and Kullback-Leibler Divergence (KLD). Denoting the truncated signature as $\mathcal{S}_{M}(\cdot)$ and the cumulative truncated signature as $\mathcal{S}^{C}_{M}(\cdot)$, the two loss functions are defined as follows:
\begin{align} \text{MSE}(\hat{S}^{\textit{fake}}_t, \hat{S}^{\textit{real}}_t)
&= \frac{1}{N}\sum_{i=1}^{N}\left(\mathcal{S}_{M}(L(\hat{S}^{\textit{fake}}_t))_i - \mathcal{S}_{M}(L(\hat{S}^{\textit{real}}_t))_i \right)^{2}   \nonumber \\
    & + \frac{1}{N}\sum_{i=1}^{N}\left(\mathcal{S}^{C}_{M}(L(\hat{S}^{\textit{fake}}_t))_i - \mathcal{S}^{C}_{M}(L(\hat{S}^{\textit{real}}_t))_i \right)^{2}\,\nonumber \\
    \text{KLD}(\hat{S}^{\textit{fake}}_t, \hat{S}^{\textit{real}}_t) &= D_{KL}\left(\mathcal{S}_{M}(L(\hat{S}^{\textit{fake}}_t))\|\mathcal{S}_{M}(L(\hat{S}^{\textit{real}}_t))) \right)  \nonumber \\
    &+D_{KL}\left(\mathcal{S}^{C}_{M}(L(\hat{S}^{\textit{fake}}_t))\|\mathcal{S}^{C}_{M}(L(\hat{S}^{\textit{real}}_t))) \right)\,, 
\label{eq:loss_functions}
\end{align}
where $N$ is the number of elements inside the $\mathcal{S}_{M}(\cdot)$ and $\mathcal{S}^{C}_{M}(\cdot)$, $D_{KL}(\cdot \| \cdot)$ is the Kullback–Leibler divergence ~\cite{kullback1951information}, and we apply the softmax function to derive probabilities from $\mathcal{S}_{M}(\cdot)$ and $\mathcal{S}^{C}_{M}(\cdot)$. We opt for conducting two distinct simulations for the model contingent on the chosen loss function.\\
Finally, the choice of LSTM as the Recurrent Block is driven by our intention to emphasize the importance of geometric patterns in time series analysis. We believe that a convincing approach to illustrate this significance is to integrate LSTM, one of the simplest recurrent neural network models in terms of complexity and architecture, with the GNN. Our goal is to show that this combination can outperform baseline models in generating synthetic time series data, even when the baseline models use more sophisticated time series models, such as the TCN used in the QuantGAN model \cite{wiese2020quant}.

%%%%%%%%%%%%%%%%%%%%%%%%%%%%%%%%%%%%%%%%%%%%%%%%%%%%%%%%%%%%%%%%%%%%%%%%
\section{Experimental Evaluation}
\label{Section_6}
In this section, we outline the approach used to compare the Sig-Graph GAN model with the baseline models. Our evaluation commences by providing a comprehensive overview of the dataset used, delineating the steps encompassing data pre-processing. Subsequently, we introduce the array of evaluation metrics adopted to facilitate rigorous comparisons. Following this, we proceed to define the hyperparameters intrinsic to both the primary model and the baseline models.\\
In the course of our analysis, the computations were executed utilizing the Nvidia GPU A100-SXM4-40GB. The optimization of hyperparameters was undertaken through the utilization of the Optuna Python package ~\cite{akiba2019optuna}. Lastly, we establish the baseline models, which encompass the Quant GAN model\footnote{Python implementation available at \url{https://github.com/ICascha/QuantGANs-replication}.}, the GARCH$(1,1)$ model ~\cite{bollerslev1986generalized}, as well as the Monte-Carlo simulation for the Black and Scholes model. 

\subsection{Dataset and Pre-processing}
For the scope of our analysis, we have selected three stock exchanges of considerable importance: the Standard \& Poor's 500 (S\&P 500), the Nasdaq Composite Index (IXIC), and the Nikkei 225 (N225). We have collected the closing prices of these stock exchanges spanning the interval from January 4, 2010, to December 30, 2019. Each dataset consists of approximately $2515$ observations.\\
Before subjecting the dataset to normalization processes to achieve a mean of zero and a variance of one, a preliminary step involves computing logarithmic returns denoted as $r_t = \log(s_t) - \log(s_{t-1})$. The choice of logarithm returns derives from their log-normal distribution characteristics. Subsequently, to induce a distribution with fat tails, we adopt the inverse Lambert-$W$ probability transform ~\cite{goerg2015lambert} as outlined in ~\cite{wiese2020quant}. 

\subsection{Evaluation Metrics}
Empirical observations in stock returns, known as stylized facts ~\cite{cont2001empirical}, demonstrated certain characteristic properties. These include fat-tailed distributions, deviating from normal distributions, as well as volatility clustering, indicating alternating periods of high and low price-change activity in historical asset returns. %Additionally, 
The presence of a leverage effect suggests a negative correlation between volatility and returns.\\
To facilitate meaningful comparisons among different models, we employ different evaluation metrics. Notably, we utilize the leverage effect score as defined in ~\cite{wiese2020quant}, and a distribution-based metric. In particular, we consider the Earth Mover's Distance (EMD), also known as the Wasserstein 1-distance, as elaborated in ~\cite{villani2009optimal,rubner2000earth}. This metric quantifies the minimal cost required to transform the distribution of real data into that of generated data. Furthermore, we calculate the Root Mean Square Error (RMSE) between the signatures of the real and generated data.

\subsection{Configurations of the Models}
\begin{table}[t]
\caption{Definition of the hyperparameters.}
\centering
\resizebox{\columnwidth}{!}{
\begin{tabular}{|l|l|l|l|l|l|l|l|}
\hline
    Loss & BS & LR & \# Neurons & \# Layers & Dropout & Seq. Len. & Dir \\
    \hline
    MSE & $30$ & $0.000797$ & $[190, 120, 190]$ & $[3,7]$ & $0.31$ & $100$ & None \\ 
    \hline
    KLD & $30$ & $0.000221$ & $[110, 70, 190]$ & $[2,4]$ & $0.35$ & $100$ & None \\
    \hline
\end{tabular}}
\label{tab:hyperparameters_result}
\end{table} % Hyperparameters table

To train our model effectively, we must define hyperparameters. These include the batch size (BS), learning rate (LR) for optimization, number of neurons (\# Neurons) for the recurrent and geometric blocks - outlined in Section \ref{Section_4} -, number of layers (\# Layers) in GNN and LSTM, dropout rate, direction (Dir) for applying the visibility graph algorithm, and sequence length (Seq. Len.) for graph derivation. We utilize RMSProp ~\cite{hinton2012neural} as our optimization algorithm. Depending on the loss function - defined in Equation (\ref{eq:loss_functions}) -, we conduct two hyperparameter optimizations. We set the number of training epochs to a fixed value of 100. Detailed results regarding the outcomes of the hyperparameter optimization process are shown in Table \ref{tab:hyperparameters_result}. The numbers of neurons respectively correspond to the GNN and LSTM within the geometric block, and the LSTM within the recurrent block, while the numbers of layers respectively correspond to the GNN and LSTM layers within the recurrent block. Finally, we consider undirected graphs. \\
For the baseline model, we adopt the parameters outlined in ~\cite{wiese2020quant} for Quant GAN. For the Monte-Carlo simulation of the geometric Brownian motion, which plays a significant role in financial mathematics for option pricing and stock price simulation. The mean and variance were determined by maximizing the log-likelihood function ~\cite{hilpisch2015derivatives}. 

\subsection{Numerical Results}
\begin{table*}[t]
\caption{Results for the real and the generated data for the Nasdaq (IXIC), Nikkei$225$ (N$225$), and Standard \& Poor's $500$ (S\&P $500$) datasets. To facilitate results comparison, the values are multiplied %scaled 
by a factor of $100$. In brackets, in the Sig-Graph GAN model is reported the type of loss function used.}
\centering
\resizebox{0.99\textwidth}{!}{\begin{tabular}
{|l||ccc||ccc||ccc||ccc||ccc|}
\hline
   {Evaluation metric}             & \multicolumn{3}{c|}{QuantGAN}                                            & \multicolumn{3}{c|}{Sig-Graph GAN(MSE)}                                        & \multicolumn{3}{c|}{Sig-Graph GAN(KLD)}                                        & \multicolumn{3}{c|}{MC}                                                                         & \multicolumn{3}{c|}{GARCH(1,1)}                                               \\ \hline
                & \multicolumn{1}{c|}{IXIC}     & \multicolumn{1}{c|}{S\&P 500}    & N225     & \multicolumn{1}{c|}{IXIC}     & \multicolumn{1}{c|}{S\&P 500}    & N225     & \multicolumn{1}{c|}{IXIC}     & \multicolumn{1}{c|}{S\& P500}    & N225     & \multicolumn{1}{c|}{IXIC}      & \multicolumn{1}{c|}{S\&P 500}     & N225                          & \multicolumn{1}{c|}{IXIC}       & \multicolumn{1}{c|}{S\& P500}     & N225       \\ \hline
EMD(1)          & \multicolumn{1}{c|}{$\mathbf{0.1323}$} & \multicolumn{1}{c|}{$0.1483$} & $0.1759$ & \multicolumn{1}{c|}{$0.1649$} & \multicolumn{1}{c|}{$\mathbf{0.1274}$} & $0.1809$ & \multicolumn{1}{c|}{$0.1618$} & \multicolumn{1}{c|}{$0.1472$} & $\mathbf{0.1611}$ & \multicolumn{1}{c|}{$64.1325$}  & \multicolumn{1}{c|}{$50.3969$}  & {$40.1514$} & \multicolumn{1}{c|}{$79.6734$}   & \multicolumn{1}{r|}{$68.8657$}  & $101.6870$   \\ \hline
EMD(5)          & \multicolumn{1}{c|}{$0.4051$} & \multicolumn{1}{c|}{$0.3681$} & $0.4341$ & \multicolumn{1}{c|}{$0.4949$} & \multicolumn{1}{c|}{$0.3144$} & $0.4232$ & \multicolumn{1}{c|}{$\mathbf{0.3931}$} & \multicolumn{1}{c|}{$\mathbf{0.2816}$} & $\mathbf{0.2066}$ & \multicolumn{1}{c|}{$321.0774$}  & \multicolumn{1}{c|}{$252.3970$}  & $203.1317$ & \multicolumn{1}{c|}{$175.3732$}   & \multicolumn{1}{c|}{$151.0271$}  & $223.6870$   \\ \hline
EMD(20)         & \multicolumn{1}{c|}{$1.4059$} & \multicolumn{1}{c|}{$1.0954$} & $1.0522$ & \multicolumn{1}{c|}{$1.3431$} & \multicolumn{1}{c|}{$\mathbf{0.8679}$} & $1.4692$ & \multicolumn{1}{c|}{$\mathbf{1.3247}$} & \multicolumn{1}{c|}{$0.9438$} & $\mathbf{0.6767}$ & \multicolumn{1}{c|}{$1284.3564$} & \multicolumn{1}{c|}{$1009.8142$} & $815.9213$ & \multicolumn{1}{c|}{$354.0737$}   & \multicolumn{1}{c|}{$305.4143$}  & $450.6941$   \\ \hline
EMD(100)        & \multicolumn{1}{c|}{$5.8784$} & \multicolumn{1}{c|}{$4.2506$} & $4.8196$ & \multicolumn{1}{c|}{$\mathbf{5.4952}$} & \multicolumn{1}{c|}{$\mathbf{4.0865}$} & $4.1957$ & \multicolumn{1}{c|}{$5.6074$} & \multicolumn{1}{c|}{$4.3141$} & $\mathbf{3.3477}$ & \multicolumn{1}{c|}{$6400.7781$} & \multicolumn{1}{c|}{$5034.7390$} & $4077.6804$ & \multicolumn{1}{c|}{$799.9104$}   & \multicolumn{1}{r|}{$695.6773$}  & $1016.9772$  \\ \hline
Sig-RMSE(1)     & \multicolumn{1}{c|}{$5.4381$} & \multicolumn{1}{c|}{$4.1200$} & $8.1107$ & \multicolumn{1}{c|}{$5.4062$} & \multicolumn{1}{c|}{$4.0911$} & $\mathbf{8.0068}$ & \multicolumn{1}{c|}{$5.3985$} & \multicolumn{1}{c|}{$\mathbf{4.0782}$} & $8.0583$ & \multicolumn{1}{c|}{$\mathbf{5.3843}$}  & \multicolumn{1}{c|}{$4.0809$}  & $8.0334$ & \multicolumn{1}{c|}{$10490.6999$} & \multicolumn{1}{c|}{$5618.0861$} & $23154.0991$ \\ \hline
Sig-RMSE(5)     & \multicolumn{1}{c|}{$5.6048$} & \multicolumn{1}{c|}{$4.2529$} & $7.7411$ & \multicolumn{1}{c|}{$5.5681$} & \multicolumn{1}{c|}{$4.2228$} & $\mathbf{7.6333}$ & \multicolumn{1}{c|}{$5.5703$} & \multicolumn{1}{c|}{$\mathbf{4.2117}$} & $7.6854$ & \multicolumn{1}{c|}{$\mathbf{5.5528}$}  & \multicolumn{1}{c|}{$4.2123$}  & $7.6637$ & \multicolumn{1}{c|}{$12185.4361$} & \multicolumn{1}{c|}{$5087.6136$} & $25530.0890$ \\ \hline
Sig-RMSE(20)    & \multicolumn{1}{c|}{$5.4281$} & \multicolumn{1}{c|}{$3.9921$} & $7.7616$ & \multicolumn{1}{c|}{$5.3587$} & \multicolumn{1}{c|}{$3.9763$} & $7.6707$ & \multicolumn{1}{c|}{$\mathbf{5.3450}$} & \multicolumn{1}{c|}{$3.9288$} & $\mathbf{7.7020}$ & \multicolumn{1}{c|}{$5.4102$}  & \multicolumn{1}{c|}{$\mathbf{3.9120}$}  & $7.6722$ & \multicolumn{1}{c|}{$15932.3498$} & \multicolumn{1}{c|}{$6263.2784$} & $24719.1718$ \\ \hline
Sig-RMSE(100)   & \multicolumn{1}{c|}{$5.5419$} & \multicolumn{1}{c|}{$4.6748$} & $8.4680$ & \multicolumn{1}{c|}{$5.4419$} & \multicolumn{1}{c|}{$4.6810$} & $8.4262$ & \multicolumn{1}{c|}{$5.4224$} & \multicolumn{1}{c|}{$4.6422$} & $8.4639$ & \multicolumn{1}{c|}{$\mathbf{5.3526}$}  & \multicolumn{1}{c|}{$\mathbf{4.5633}$}  & $\mathbf{8.3349}$ & \multicolumn{1}{c|}{$22124.6168$} & \multicolumn{1}{c|}{$6461.6048$} & $54772.4263$ \\ \hline
Leverage Effect & \multicolumn{1}{c|}{$\mathbf{3.8110}$} & \multicolumn{1}{c|}{$3.8231$} & $3.5620$ & \multicolumn{1}{c|}{$4.8641$} & \multicolumn{1}{c|}{$3.9510$} & $4.7748$ & \multicolumn{1}{c|}{$3.9694$} & \multicolumn{1}{c|}{$\mathbf{3.6354}$} & $\mathbf{3.0600}$ & \multicolumn{1}{c|}{$4.1356$}  & \multicolumn{1}{c|}{$4.125$}   & $3.5627$ & \multicolumn{1}{c|}{$3.8905$}   & \multicolumn{1}{c|}{$3.9218$}  & $3.4887$  \\ \hline
\end{tabular}}
\label{tab:results}
\end{table*} % Results table
Data generation is conducted at different temporal intervals: daily, weekly, monthly, and long-term, corresponding to $1$, $5$, $20$, and $100$ days, respectively. Results for both real and generated data for the various datasets are presented in Table \ref{tab:results}. Optimal results are highlighted in bold.\\
Our proposed model consistently outperforms the baseline models in terms of the EMD and leverage effect metric. The choice between KLD and MSE loss depends on the dataset, favoring KLD for N225 and IXIC datasets, while MSE loss for S\&P $500$. However, the Quant GAN model excels for the IXIC dataset in both one-day EMD and the leverage effect metric. The Sig-MSE metric outcomes show a balance between our model and the Monte Carlo simulation for geometric Brownian motion. This balance is due to the Monte Carlo approach solely relying on the mean and standard deviation for stock price trajectory simulation.\\
\begin{figure}[ht]
\centering
\includegraphics[width=1\columnwidth, height=0.30\textheight]{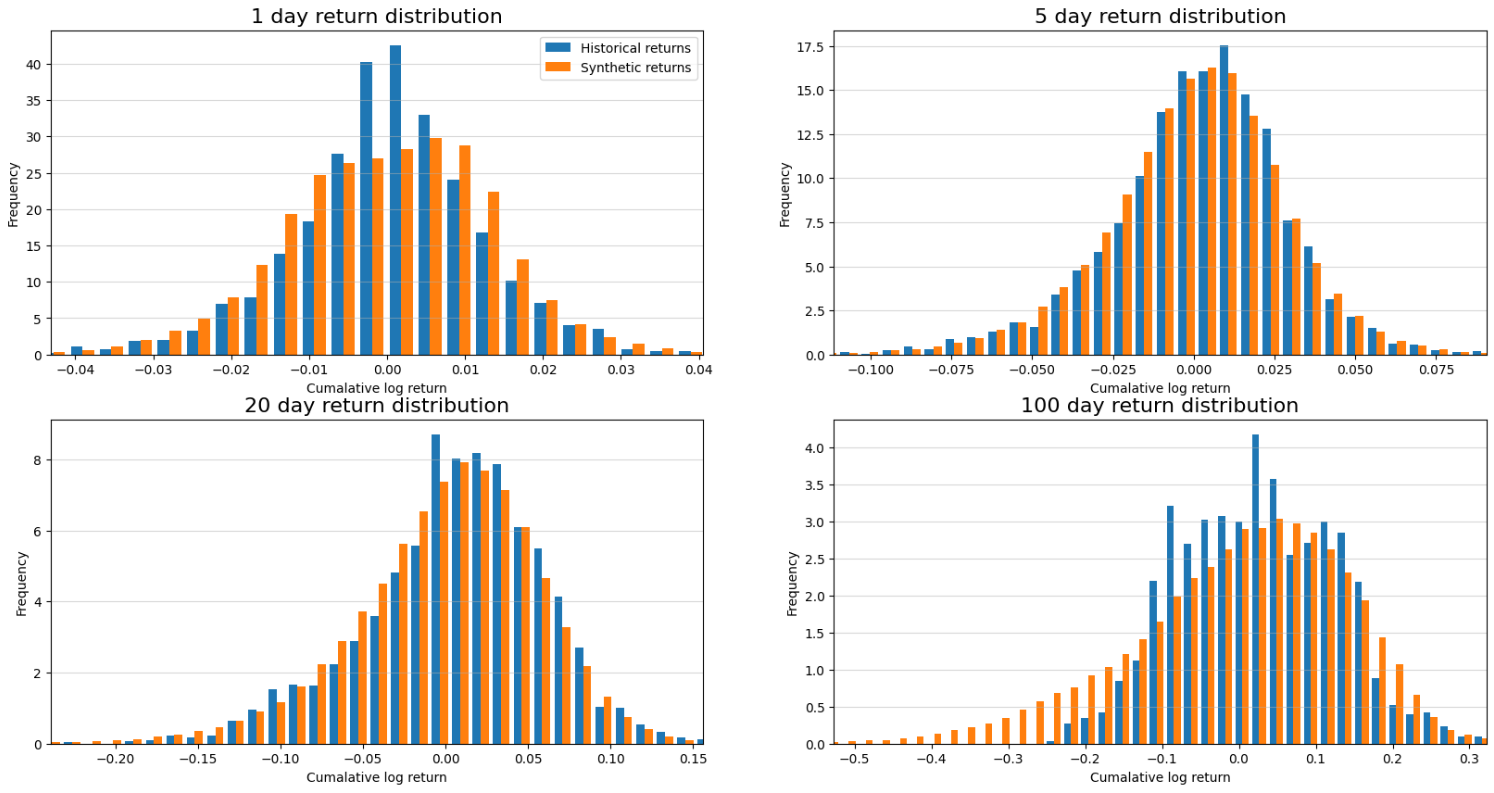}
\caption{Sig-Graph GAN (KLD) Cumulative Returns Distribution of the N225 for the historical returns (blue) and for the synthetic returns (orange).}
\label{fig: cum_ret} 
\end{figure}
Figure \ref{fig: cum_ret} shows the cumulative log-returns distributions from the Sig-Graph GAN model using the KLD loss on the N$225$ dataset. The model effectively replicates the distributions for $5$ and $10$ days, showing similar results. For the $1$-day scenario, it captures the tail characteristics but not the high portion of the curve.

\subsection{Ablation Study}
\begin{figure}[ht]
\centering
\includegraphics[width=1\columnwidth, height=0.22\textheight]{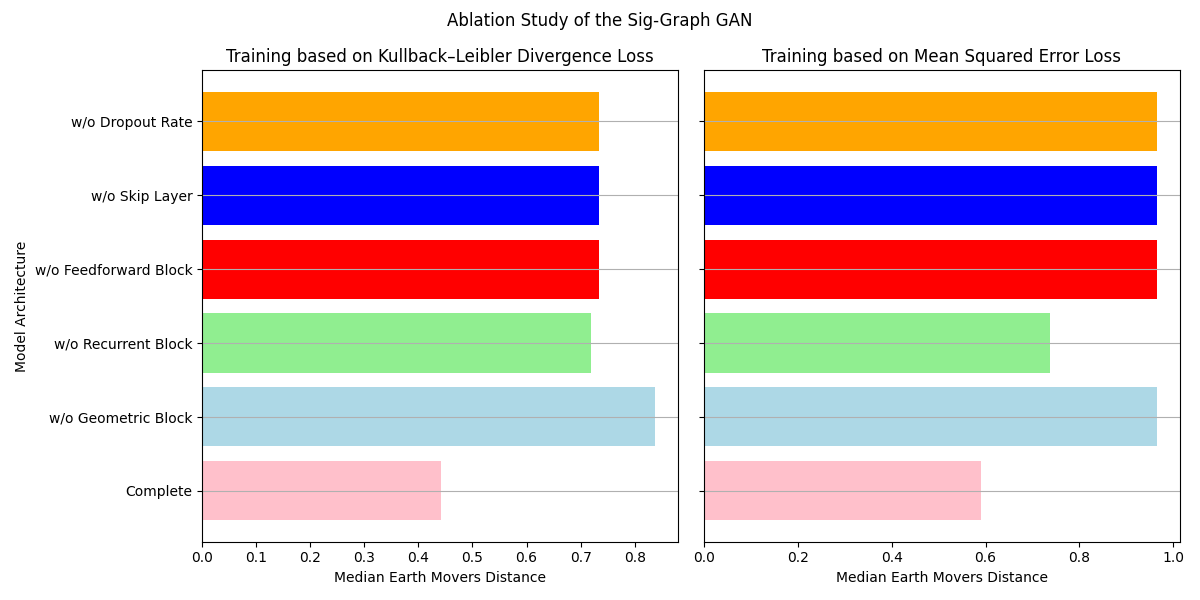}
\caption{An ablation study was conducted for the Sig-Graph GAN trained using the two custom loss functions. The first plot refers to the Sig-Graph GAN (KLD) trained on the Nikkei $225$ data, while the second plot refers to the Sig-Graph GAN (MSE) trained on the S\&P $500$.}
\label{fig:ablation_study} 
\end{figure}

To evaluate the importance of the different components in the Sig-Graph GAN model, we train the model by removing one component at a time while keeping all others fixed. The main components considered in the proposed model are the Geometric Block, Recurrent Block, Feedforward Block, skip layer, and dropout rate. Figure \ref{fig:ablation_study} presents the results of this ablation study, for the proposed model trained with the two custom loss functions on the N$225$ and S\&P $500$ datasets, depicted in the first and second figures, respectively. Due to space limitations, we evaluated the performance using only the (widely adopted) EMD evaluation metric, denoting models obtained by removing specific components, for exanple, as ``w/o Geometric Block''.  
Lower values and smaller box sizes indicate better results. We observe that the importance of the components varies depending on the type of loss function used to train the model. Specifically, we find that without the Geometric Block, the model performs worse with both loss functions, with the impact of the Geometric Block more evident with the KLD loss. It is interesting to note that the contribution of the dropout rate, skip layer, and Feedforward Block remain consistent across both loss functions, with the impact and importance of this component more evident when the MSE loss is used. 
%However, the contributions of the dropout rate, skip layer, and Feedforward Block remain consistent across both loss functions. Specifically, we find that without the Geometric Block, the model performs worse when the KLD loss is used, while without the Recurrent Block, it performs better when the MSE loss is used.

%%%%%%%%%%%%%%%%%%%%%%%%%%%%%%%%%%%%%%%%%%%%%%%%%%%%%%%%%%%%%%%%%%%%%%%%
\section{Conclusion}
\label{Section_7}
In this study, we introduce a novel approach that combines GNN, LSTM networks, and Signature transformation to construct a GAN model for the generation of synthetic stock log-returns. Our methodology leverages the inherent geometric patterns present within the time series data, leading to enhanced capabilities in the synthesis of artificial data. Through comprehensive evaluations conducted on multiple datasets: IXIC, N$225$, and S\&P $500$. We demonstrate that our proposed model consistently surpasses baseline models, as indicated by improved performance in metrics such as EMD and leverage effect. While the Sig-MSE metric does not always yield the best results, our model showcases competitive outcomes.
Future works could involve extending Sig-Graph GAN model to tackle other time series generation challenges, as well as assessing its potential in enhancing the performance of trading strategies based on synthetic data. Furthermore, future research endeavors will also explore the possibility of replacing the LSTM within the Recurrent Block with more sophisticated time series models, such as Transformers.

%%%%%%%%%%%%%%%%%%%%%%%%%%%%%%%%%%%%%%%%%%%%%%%%%%%%%%%%%%%%%%%%%%%%%%%%

%%% Use this environment to include acknowledgements (optional).
%%% This will be omitted in doubleblind mode.

%\begin{ack}
%By using the \texttt{ack} environment to insert your (optional) 
%acknowledgements, you can ensure that the text is suppressed whenever 
%you use the \texttt{doubleblind} option. In the final version, 
%acknowledgements may be included on the extra page intended for references.
%\end{ack}

%%%%%%%%%%%%%%%%%%%%%%%%%%%%%%%%%%%%%%%%%%%%%%%%%%%%%%%%%%%%%%%%%%%%%%%%

%%% Use this command to include your bibliography file.
\bibliography{mybibfile}

@book{Box2015,
  author={Box, George EP and Jenkins, Gwilym M and Reinsel, Gregory C and Ljung, Greta M},
  title={Time Series Analysis: Forecasting and Control},
  year={2015},
  publisher={John Wiley \& Sons},
  edition={5}
}

@book{Tsay2005,
  title={Analysis of Financial Time Series},
  author={Tsay, Ruey S},
  year={2005},
  publisher={John Wiley \& Sons},
}

@article{Black1973pricing,
  title={The pricing of options and corporate liabilities},
  author={Black, Fischer and Scholes, Myron},
  journal={Journal of Political Economy},
  volume={81},
  number={3},
  pages={637--654},
  year={1973},
  publisher={The University of Chicago Press}
}

@article{heston1993closed,
  title={A closed-form solution for options with stochastic volatility with applications to bond and currency options},
  author={Heston, Steven L},
  journal={The Review of Financial Studies},
  volume={6},
  number={2},
  pages={327--343},
  year={1993},
  publisher={Oxford University Press}
}

@book{lamberton2011introduction,
  title={Introduction to Stochastic Calculus Applied to Finance},
  author={Lamberton, Damien and Lapeyre, Bernard},
  year={2011},
  publisher={CRC press}
}

@book{meucci2005risk,
  title={Risk and Asset Allocation},
  author={Meucci, Attilio},
  volume={1},
  year={2005},
  publisher={Springer}
}

@book{buff2002uncertain,
  title={Uncertain Volatility Models: Theory and Application},
  author={Buff, Robert},
  year={2002},
  publisher={Springer Science \& Business Media}
}

@book{peters1994fractal,
  title={Fractal Market Analysis: Applying Chaos Theory to Investment and Economics},
  author={Peters, Edgar E},
  volume={24},
  year={1994},
  publisher={John Wiley \& Sons}
}

@article{evertsz1995fractal,
  title={Fractal Geometry of Financial Time Series},
  author={Evertsz, Carl JG},
  journal={Fractals},
  volume={3},
  number={03},
  pages={609--616},
  year={1995},
  publisher={World Scientific}
}

@article{goodfellow2020generative,
  title={Generative adversarial networks},
  author={Goodfellow, Ian and Pouget-Abadie, Jean and Mirza, Mehdi and Xu, Bing and Warde-Farley, David and Ozair, Sherjil and Courville, Aaron and Bengio, Yoshua},
  journal={Communications of the ACM},
  volume={63},
  number={11},
  pages={139--144},
  year={2020},
  publisher={ACM New York, NY, USA}
}

@article{lacasa2008time,
  title={From time series to complex networks: The visibility graph},
  author={Lacasa, Lucas and Luque, Bartolo and Ballesteros, Fernando and Luque, Jordi and Nuno, Juan Carlos},
  journal={Proceedings of the National Academy of Sciences},
  volume={105},
  number={13},
  pages={4972--4975},
  year={2008},
  publisher={National Acad Sciences}
}

@book{mandelbrot2013fractals,
  title={Fractals and Scaling in Finance: Discontinuity, Concentration, Risk. Selecta volume E},
  author={Mandelbrot, Benoit B},
  year={2013},
  publisher={Springer Science \& Business Media}
}

@article{scarselli2008graph,
  title={The graph neural network model},
  author={Scarselli, Franco and Gori, Marco and Tsoi, Ah Chung and Hagenbuchner, Markus and Monfardini, Gabriele},
  journal={IEEE Transactions on Neural Networks},
  volume={20},
  number={1},
  pages={61--80},
  year={2008},
  publisher={IEEE}
}

@article{lyons1998differential,
  title={Differential equations driven by rough signals},
  author={Lyons, Terry J},
  journal={Revista Matem{\'a}tica Iberoamericana},
  volume={14},
  number={2},
  pages={215--310},
  year={1998}
}

@inproceedings{lyons2014feature,
  title={A feature set for streams and an application to high-frequency financial tick data},
  author={Lyons, Terry and Ni, Hao and Oberhauser, Harald},
  booktitle={Proceedings of the 2014 International Conference on Big Data Science and Computing},
  pages={1--8},
  year={2014}
}

@book{resnick2019probability,
  title={A {P}robability {P}ath},
  author={Resnick, Sidney},
  year={2019},
  publisher={Springer}
}

@article{hochreiter1997long,
  title={Long short-term memory},
  author={Hochreiter, Sepp and Schmidhuber, J{\"u}rgen},
  journal={Neural Computation},
  volume={9},
  number={8},
  pages={1735--1780},
  year={1997},
  publisher={MIT press}
}

@article{fama1970efficient,
  title={Efficient capital markets: A review of theory and empirical work},
  author={Fama, Eugene F},
  journal={The Journal of Finance},
  volume={25},
  number={2},
  pages={383--417},
  year={1970},
  publisher={JSTOR}
}

@book{brockwell2002introduction,
  title={Introduction to Time Series and Forecasting},
  author={Brockwell, Peter J and Davis, Richard A},
  year={2002},
  publisher={Springer}
}

@article{bollerslev1986generalized,
  title={Generalized autoregressive conditional heteroskedasticity},
  author={Bollerslev, Tim},
  journal={Journal of Econometrics},
  volume={31},
  number={3},
  pages={307--327},
  year={1986},
  publisher={Elsevier}
}

@article{engle2012arch,
  title={{ARCH}/{GARCH} models in applied financial econometrics},
  author={Engle, Robert F and Focardi, Sergio M and Fabozzi, Frank J},
  journal={Encyclopedia of Financial Models},
  year={2012},
  publisher={Wiley Online Library}
}

@article{ruiz2002bootstrapping,
  title={Bootstrapping Financial Time Series},
  author={Ruiz, Esther and Pascual, Lorenzo},
  journal={Journal of Economic Surveys},
  volume={16},
  number={3},
  pages={271--300},
  year={2002},
  publisher={Wiley Online Library}
}

@article{merton1976option,
  title={Option pricing when underlying stock returns are discontinuous},
  author={Merton, Robert C},
  journal={Journal of Financial Economics},
  volume={3},
  number={1-2},
  pages={125--144},
  year={1976},
  publisher={Elsevier}
}

@article{kou2002jump,
  title={A jump-diffusion model for option pricing},
  author={Kou, Steven G},
  journal={Management Science},
  volume={48},
  number={8},
  pages={1086--1101},
  year={2002},
  publisher={INFORMS}
}

@book{wang2012monte,
  title={Monte Carlo Simulation with Applications to Finance},
  author={Wang, Hui},
  year={2012},
  publisher={CRC Press}
}

@article{kondratyev2019market,
  title={The market generator},
  author={Kondratyev, Alexei and Schwarz, Christian},
  journal={Available at SSRN 3384948},
  year={2019}
}

@article{de2019enriching,
  title={Enriching financial datasets with generative adversarial networks},
  author={De Meer Pardo, Fernando},
  journal={MS thesis, Delft University of Technology, The Netherlands},
  year={2019}
}

@misc{lyons2014rough,
  title={Rough paths, signatures and the modelling of functions on streams},
  author={Terry Lyons},
  year={2014},
  eprint={1405.4537},
  archivePrefix={arXiv},
  primaryClass={math.PR}
}

@article{wiese2020quant,
  title={Quant {GAN}s: Deep generation of financial time series},
  author={Wiese, Magnus and Knobloch, Robert and Korn, Ralf and Kretschmer, Peter},
  journal={Quantitative Finance},
  volume={20},
  number={9},
  pages={1419--1440},
  year={2020},
  publisher={Taylor \& Francis}
}

@misc{ni2020conditional,
  title={Conditional sig-{W}asserstein {GAN}s for time series generation},
  author={Hao Ni and Lukasz Szpruch and Magnus Wiese and Shujian Liao and Baoren Xiao},
  year={2020},
  eprint={2006.05421},
  archivePrefix={arXiv},
  primaryClass={cs.LG}
}

@article{de2022tackling,
  title={Tackling the exponential scaling of signature-based generative adversarial networks for high-dimensional financial time-series generation},
  author={De Meer Pardo, Fernando and Schwendner, Peter and Wunsch, Marcus},
  journal={The Journal of Financial Data Science},
  volume={4},
  number={4},
  pages={110--132},
  year={2022},
  publisher={Portfolio Management Research}
}

@misc{chevyrev2016primer,
  title={A primer on the signature method in machine learning},
  author={Ilya Chevyrev and Andrey Kormilitzin},
  year={2016},
  eprint={1603.03788},
  archivePrefix={arXiv},
  primaryClass={stat.ML}
}

@article{wang2019time,
  title={Time-dependent graphs: Definitions, applications, and algorithms},
  author={Wang, Yishu and Yuan, Ye and Ma, Yuliang and Wang, Guoren},
  journal={Data Science and Engineering},
  volume={4},
  pages={352--366},
  year={2019},
  publisher={Springer}
}

@article{stephen2015visibility,
  title={Visibility Graph Based Time Series Analysis},
  author={Stephen, Mutua and Gu, Changgui and Yang, Huijie},
  journal={PloS One},
  volume={10},
  number={11},
  pages={e0143015},
  year={2015},
  publisher={Public Library of Science San Francisco, CA USA}
}

@misc{Bergillos2020visibility,
  title        = "Ts2vg: Time series to visibility graphs",
  author       = {Carlos Bergillos},
  howpublished = "\url{https://pypi.org/project/ts2vg/}",
  year         = 2020,
  note         = "Accessed: 2028-08-09"
}

@inproceedings{gilmer2017neural,
  title={Neural message passing for quantum chemistry},
  author={Gilmer, Justin and Schoenholz, Samuel S and Riley, Patrick F and Vinyals, Oriol and Dahl, George E},
  booktitle={International Conference on Machine Learning},
  pages={1263--1272},
  year={2017},
  organization={PMLR}
}

@inproceedings{xu2018representation,
  title={Representation learning on graphs with jumping knowledge networks},
  author={Xu, Keyulu and Li, Chengtao and Tian, Yonglong and Sonobe, Tomohiro and Kawarabayashi, Ken-ichi and Jegelka, Stefanie},
  booktitle={International Conference on Machine Learning},
  pages={5453--5462},
  year={2018},
  organization={PMLR}
}

@article{kullback1951information,
  title={On Information and Sufficiency},
  author={Kullback, Solomon and Leibler, Richard A},
  journal={The Annals of Mathematical Statistics},
  volume={22},
  number={1},
  pages={79--86},
  year={1951},
  publisher={JSTOR}
}

@inproceedings{you2022roland,
  title={{ROLAND}: Graph learning framework for dynamic graphs},
  author={You, Jiaxuan and Du, Tianyu and Leskovec, Jure},
  booktitle={Proceedings of the 28th ACM SIGKDD Conference on Knowledge Discovery and Data Mining},
  pages={2358--2366},
  year={2022}
}

@misc{levin2013learning,
  title={Learning from the past, predicting the statistics for the future, learning an evolving system},
  author={Levin, Daniel and Lyons, Terry and Ni, Hao},
  year={2013},
  eprint={1309.0260},
  archivePrefix={arXiv},
  primaryClass={q-fin.ST}
}

@article{chen1958integration,
  title={Integration of paths -- A faithful representation of paths by noncommutative formal power series},
  author={Chen, Kuo-Tsai},
  journal={Transactions of the American Mathematical Society},
  volume={89},
  number={2},
  pages={395--407},
  year={1958},
  publisher={JSTOR}
}

@inproceedings{lemercier2021distribution,
  title={Distribution regression for sequential data},
  author={Lemercier, Maud and Salvi, Cristopher and Damoulas, Theodoros and Bonilla, Edwin and Lyons, Terry},
  booktitle={International Conference on Artificial Intelligence and Statistics},
  pages={3754--3762},
  year={2021},
  organization={PMLR}
}

@article{chevyrev2016characteristic,
  title={Characteristic functions of measures on geometric rough paths},
  author={Chevyrev, Ilya and Lyons, Terry},
journal={The Annals of Probability},
  volume={44},
  number={6},
  pages={4049--4082},
  year={2016},
  publisher={Ann. Probab}
}

@misc{kipf2016semi,
  title={Semi-supervised classification with graph convolutional networks},
  author={Kipf, Thomas N and Welling, Max},
  year={2016},
  eprint={1609.02907},
  archivePrefix={arXiv},
  primaryClass={cs.LG}
}

@inproceedings{he2015delving,
  title={Delving deep into rectifiers: Surpassing human-level performance on imagenet classification},
  author={He, Kaiming and Zhang, Xiangyu and Ren, Shaoqing and Sun, Jian},
  booktitle={Proceedings of the IEEE international conference on computer vision},
  pages={1026--1034},
  year={2015}
}

@article{lyons2015expected,
  title={Expected signature of {B}rownian motion up to the first exit time from a bounded domain},
  author={Lyons, Terry and Ni, Hao},
  year={2015},
  journal={The Annals of Probability},
  volume={43},
  number={5},
  pages={2729-2762},
  publisher={Ann. Probab} 
}

@inproceedings{akiba2019optuna,
  title={Optuna: A next-generation hyperparameter optimization framework},
  author={Akiba, Takuya and Sano, Shotaro and Yanase, Toshihiko and Ohta, Takeru and Koyama, Masanori},
  booktitle={Proceedings of the 25th ACM SIGKDD International Conference on Knowledge Discovery \& Data Mining},
  pages={2623--2631},
  year={2019}
}

@article{goerg2015lambert,
  title={The {L}ambert way to {G}aussianize heavy-tailed data with the inverse of {T}ukey's h transformation as a special case},
  author={Goerg, Georg M},
  journal={The Scientific World Journal},
  volume={2015},
  year={2015},
  publisher={Hindawi Limited}
}

@article{cont2001empirical,
  title={Empirical properties of asset returns: Stylized facts and statistical issues},
  author={Cont, Rama},
  journal={Quantitative Finance},
  volume={1},
  number={2},
  pages={223},
  year={2001},
  publisher={IOP Publishing}
}

@book{villani2009optimal,
  title={Optimal Transport: Old and New},
  author={Villani, C{\'e}dric and others},
  volume={338},
  year={2009},
  publisher={Springer}
}

@article{rubner2000earth,
  title={The Earth Mover's Distance as a Metric for Image Retrieval},
  author={Rubner, Yossi and Tomasi, Carlo and Guibas, Leonidas J},
  journal={International Journal of Computer Vision},
  volume={40},
  pages={99--121},
  year={2000},
  publisher={Springer}
}

@techreport{hinton2012neural,
  title={Neural Networks for Machine Learning: Lecture 6a -- Overview of Mini-Batch Gradient Descent},
  author={Hinton, Geoffrey and Srivastava, Nitish and Swersky, Kevin},
  type    = "Technical Report",
  institution = "Dept.\ of Computer Science, University of Toronto",
  year={2012}
}

@book{hilpisch2015derivatives,
  title={Derivatives Analytics with {P}ython: Data Analysis, Models, Simulation, Calibration and Hedging},
  author={Hilpisch, Yves},
  year={2015},
  publisher={John Wiley \& Sons}
}

@misc{kidger2020signatory,
  title={Signatory: Differentiable computations of the signature and logsignature transforms, on both {CPU} and {GPU}},
  author={Kidger, Patrick and Lyons, Terry},
  year={2020},
  eprint={2001.00706},
  archivePrefix={arXiv},
  primaryClass={cs.LG}
}

@inproceedings{yu2017seqgan,
  title={Seq{GAN}: Sequence Generative Adversarial Nets with Policy Gradient},
  author={Lantao, Yu and Weinan, Zhang and Jun, Wang and Yong, Yu},
  booktitle={Proceedings of the 31st AAAI Conference on Artificial Intelligence (AAAI’17)},
  pages={2852-–2858},
  year={2017}
}

@inproceedings{mogren2016crnngan, 
  title={{C-RNN-GAN}: A continuous recurrent neural network with adversarial training}, 
  author={Olof Mogren}, 
  booktitle={Constructive Machine Learning Workshop (CML) at NIPS 2016}, 
  pages={1}, 
  year={2016} 
}

@inproceedings{Yoon2019TimeGAN,
 author = {Yoon, Jinsung and Jarrett, Daniel and van der Schaar, Mihaela},
 booktitle = {Advances in Neural Information Processing Systems},
 editor = {H. Wallach and H. Larochelle and A. Beygelzimer and F. d\textquotesingle Alch\'{e}-Buc and E. Fox and R. Garnett},
 pages = {},
 publisher = {Curran Associates, Inc.},
 title = {Time-series Generative Adversarial Networks},
 volume = {32},
 year = {2019}
}

@misc{sun2023decisionaware,
      title={Decision-Aware Conditional {GAN}s for Time Series Data}, 
      author={He Sun and Zhun Deng and Hui Chen and David C. Parkes},
      year={2023},
      eprint={2009.12682},
      archivePrefix={arXiv},
      primaryClass={cs.LG}
}

@inproceedings{assefa2020generating,
  title={Generating synthetic data in finance: opportunities, challenges and pitfalls},
  author={Assefa, Samuel A and Dervovic, Danial and Mahfouz, Mahmoud and Tillman, Robert E and Reddy, Prashant and Veloso, Manuela},
  booktitle={Proceedings of the First ACM International Conference on AI in Finance},
  pages={1--8},
  year={2020}
}

\end{document}